%% file: final.tex
\documentclass[twocolumn,english,aps,prd,reprint,floatfix,notitlepage,footinbib,preprintnumbers,superscriptaddress]{revtex4-1}
\pdfoutput=1
\usepackage{lmodern}

\usepackage[T1]{fontenc}
\usepackage[latin9]{inputenc}
\usepackage{geometry}
\usepackage{subfigure,lmodern, amsmath,amssymb, graphicx, pifont, adjustbox, bm, xcolor}
\usepackage{amsfonts}
\usepackage{geometry}
\usepackage{comment}
\usepackage{mathtools}
\usepackage{float}
\usepackage{slashed}
\usepackage{ragged2e}
\usepackage{array}
\usepackage{ulem}

\makeatletter\g@addto@macro\bfseries{\boldmath}\makeatother

\usepackage{stackengine}
\usepackage{esint}
\usepackage[unicode=true,pdfusetitle,
 bookmarks=true,bookmarksnumbered=false,bookmarksopen=false,
 breaklinks=false,pdfborder={0 0 1},backref=false,colorlinks=true]
 {hyperref}
\hypersetup{
 pdfauthor={Gary T. Horowitz, Maciej Kolanowski, Grant N. Remmen, Jorge E. Santos},
 citecolor=black,linkcolor=black,urlcolor=black}

\newcommand{\appendixref}[1]{\hyperref[#1]{appendix~\ref{#1}}}
\def\equationautorefname~#1\null{eq.\,(#1)\null}
\usepackage{breakurl}
\usepackage[hang,flushmargin]{footmisc} 
\allowdisplaybreaks
\makeatletter

\usepackage{etoolbox}
\apptocmd{\thebibliography}{\justifying\setlength{\leftskip}{7.4mm}}{}{} 
 
 \usepackage{relsize}
\usepackage{babel}

\makeatletter
\def\simgt{\mathrel{\lower2.5pt\vbox{\lineskip=0pt\baselineskip=0pt
           \hbox{$>$}\hbox{$\sim$}}}}
\def\simlt{\mathrel{\lower2.5pt\vbox{\lineskip=0pt\baselineskip=0pt
           \hbox{$<$}\hbox{$\sim$}}}}
\makeatother

\usepackage{changepage}

\newcommand{\be}{\begin{equation}}
\newcommand{\ee}{\end{equation}}
\newcommand{\bea}{\begin{eqnarray}}
\newcommand{\eea}{\end{eqnarray}}

\newcolumntype{P}[1]{>{\centering\arraybackslash}p{#1}}

\usepackage{color}
    \definecolor{darkgreen}{rgb}{0,0.5,0}
    \definecolor{darkred}{rgb}{0.5,0,0}
    \definecolor{darkblue}{rgb}{0,0,0.6}
    \definecolor{purple}{rgb}{0.4,.2,0.7}

\begin{document}

\title{Extremal Kerr black holes as amplifiers of new physics}

\author{Gary~T.~Horowitz}
\affiliation{Department of Physics, University of California, Santa Barbara, CA 93106, U.S.A.}
\author{Maciej~Kolanowski}
\affiliation{Institute of Theoretical Physics, Faculty of Physics, University of Warsaw, Pasteura 5, 02-093 Warsaw, Poland\looseness=-1}
\author{Grant~N.~Remmen}
\affiliation{Department of Physics, University of California, Santa Barbara, CA 93106, U.S.A.}
\affiliation{Kavli Institute for Theoretical Physics, University of California, Santa Barbara, CA 93106, U.S.A.}
\author{Jorge~E.~Santos}
\affiliation{Department of Applied Mathematics and Theoretical Physics, University of Cambridge, Wilberforce Road, Cambridge, CB3 0WA, UK}

\begin{abstract}
\noindent We show that extremal Kerr black holes are sensitive probes of new physics. Stringy or quantum corrections to general relativity are expected to generate higher-curvature terms in the gravitational action. We show that in the presence of these terms, 
asymptotically flat extremal rotating black holes have curvature singularities on their horizon. Furthermore, near-extremal black holes can have large yet finite tidal forces for infalling observers. 
In addition, we consider five-dimensional extremal charged black holes and show that  higher-curvature terms can have a large effect on the horizon geometry.
\end{abstract}
\maketitle
\section{Introduction}

Various arguments suggest that general relativity is a low-energy approximation to a more complete theory. When the effective action is written just in terms of the metric (and possibly other light fields),  it includes higher-curvature terms in addition to the usual Einstein-Hilbert term. This occurs whenever there are new classical high-energy degrees of freedom, such as in string theory, or when quantum corrections are included from integrating out massive states.  It is often assumed that these higher-order terms are negligible in regions of low curvature and can be ignored. We show that this is not always the case. Extremal Kerr black holes turn out to be very sensitive to these small corrections.

Although the near-horizon geometry of an extremal black hole receives only a small correction from these higher-derivative terms, generic stationary, axisymmetric perturbations generate singularities on the horizon. When one extends the near-horizon geometry to a full asymptotically flat solution, one necessarily introduces these singular perturbations.
So the full solution has a singular horizon. 
Infalling timelike or null geodesics feel infinite tidal forces crossing the horizon \footnote{This effect is different from the dynamical instability of extremal black holes discussed in  Refs.~\cite{Aretakis:2012ei,Gajic:2023uwh}.}.  This singularity is very similar to that found recently for generic extremal black holes in anti-de Sitter space \cite{Horowitz:2022mly}. As in that case, near-extremal black holes have large (finite) tidal forces, but curvature scalars always remain small. (See Ref.~\cite{Chen:2018jed} for an earlier example of higher-curvature terms leading to singular extremal horizons in a theory with a scalar dilaton.)

In a purely gravitational effective field theory (EFT)---or even in a theory containing matter, but in the case where we are interested in vacuum solutions---we can build the action as a functional of the Riemann tensor alone.
Any terms containing the Ricci tensor or scalar curvature can be removed order-by-order in the derivative expansion via a field redefinition of the graviton, e.g., $h_{ab} \rightarrow h_{ab} + c R_{ab}$.
Modulo various tensor identities~\cite{Fulling}, the only curvature invariant at fourth order in derivatives that is independent of the Ricci tensor is $R_{abcd}R^{abcd}$.
This is field redefinition equivalent to the Gauss-Bonnet term $R_{abcd}R^{abcd} - 4R_{ab}R^{ab} + R^2$, which in four spacetime dimensions is topological and therefore cannot affect the bulk equations of motion.
At the next order, there are two independent operators in four dimensions constructible from the Riemann tensor,
$\nabla_a R_{bcde}\nabla^a R^{bcde}$ and $R_{ab}^{\phantom{ab}cd}R_{cd}^{\phantom{cd}ef}R_{ef}^{\phantom{ef}ab}$. 
The former can be rearranged, up to a total derivative, into a Riemann-cubed term and operators involving the Ricci tensor, so we can drop it without loss of generality. Similarly, there are two independent $(R_{abcd})^4$ terms \footnote{We are only counting parity-even terms, since one can show that parity-odd terms do not affect our result.}.

We are thus left with the leading effective theory~\cite{Endlich:2017tqa}
\begin{equation}
{\cal L} = \frac{1}{2\kappa^2}\left(R + \eta\,\kappa^4 {\cal R}^3 +\lambda\,\kappa^6\mathcal{C}^2 + \tilde{\lambda}\,\kappa^6\tilde{\mathcal{C}}^2\right),
\label{eq:action}
\end{equation}
where we write Newton's constant in terms of $\kappa^2=8\pi G$ and define ${\cal R}^3\equiv R_{ab}^{\phantom{ab}cd}R_{cd}^{\phantom{cd}ef}R_{ef}^{\phantom{ef}ab}$, $\mathcal{C}\equiv R_{abcd}R^{abcd}$, and $\tilde{\mathcal{C}}\equiv \tilde{R}_{abcd}R^{abcd}$,
with the dual tensor $\tilde{R}_{abcd}=\epsilon_{ab}^{\phantom{ab}pq}R_{pqcd}$~\footnote{We work with the sign conventions $\epsilon_{0123}=\sqrt{-g}$, $R_{ab} = R^c_{\;\;acb}$, $R^a_{\;\;bcd} = \partial_c \Gamma^a_{bd} - \partial_d \Gamma^a_{bc} + \Gamma^a_{ce}\Gamma^e_{bd} - \Gamma^a_{de}\Gamma^e_{bc}$, $\Gamma^a_{bc} = g^{ad}(\partial_c g_{db} + \partial_b g_{dc} - \partial_d g_{bc})/2$, and mostly-plus metric signature}. In the above, $\eta$, $\lambda$, and $\tilde{\lambda}$ are dimensionless.

The equations of motion that follow from Eq.~\eqref{eq:action}, around background solutions satisfying $R_{ab}\,{=}\,0$, can be written as \cite{Endlich:2017tqa,Cardoso:2018ptl,Cano:2019ore} 
\begin{equation}
R_{ab}-\frac{1} {2}R g_{ab}=T^{\rm cubic}_{ab}+T^{\rm quartic}_{ab}\label{eq:EOMpre}
\end{equation}
with
\begin{widetext}
\begin{equation}
\begin{aligned}
T^{\rm cubic}_{ab}&=\eta\,\kappa^4\,\left[3\,R_{a}^{\phantom{a}cde}R_{de}^{\phantom{de}gh}R_{ghcb}+\frac{1}{2}g_{ab}R_{gh}^{\phantom{gh}cd}R_{cd}^{\phantom{cd}ef}R_{ef}^{\phantom{ef}gh}-6 \nabla^c\nabla^d\left(R_{acgh}R_{bd}^{\phantom{bd}gh}\right)\right]
\\
T^{\rm quartic}_{ab} & =-\lambda\,\kappa^6\,\left(8 R_{acbd}\nabla^c\nabla^d \mathcal{C}+\frac{g_{ab}}{2} \mathcal{C}^2\right)-\tilde{\lambda}\,\kappa^6\,\left(8 \tilde{R}_{acbd}\nabla^c\nabla^d \tilde{\mathcal{C}}+\frac{g_{ab}}{2}\tilde{\mathcal{C}}^2\right),
\end{aligned}
\label{eq:EOM}
\end{equation}
\end{widetext}
which should be regarded as providing the leading corrections to vacuum solutions to linear order in $\eta$, $\lambda$, and $\tilde{\lambda}$; see also Ref.~\cite{Reall:2019sah}.
We will be interested in computing how these corrections affect, in a surprisingly dramatic fashion, the behavior of perturbations to extremal black holes.

\section{Extremal Kerr}
We expand the metric as
\begin{equation}
g_{ab}=g^{(0)}_{ab}+\eta\,h^{(6)}_{ab}+\lambda\,h^{(8)}_{ab}+\tilde{\lambda}\,\tilde{h}^{(8)}_{ab},
\end{equation}
with $g^{(0)}$ satisfying the vacuum Einstein equation, and solve Eqs.~\eqref{eq:EOMpre} and \eqref{eq:EOM} to linear order in the EFT coefficients $\{\eta,\lambda,\tilde{\lambda}\}$~\footnote{We note that the second-order perturbation $\delta h^{(6)}_{ab}$ from ${\cal R}^3$ at $O(\eta^2)$ is smaller than the linear-order quartic-Riemann perturbation $h_{ab}^{(8)}$ via the following reasoning. The linear perturbation from the cubic-Riemann term, by Eqs.~\eqref{eq:EOMpre} and \eqref{eq:EOM}, is given schematically by $\Box h^{(6)}_{ab} \sim \eta {\cal R}^3$, so $h^{(6)}_{ab} \sim \eta \kappa^4/J^2$. Then the $O(\eta^2)$ backreaction effect is generated via the equation of motion as $\Box \delta h^{(6)}_{ab} \sim (\nabla h_{ab}^{(6)})^2  + \eta {\cal R}^2 \delta {\cal R} \sim \eta^2 \kappa^8/J^5$, so $\delta h^{(6)}_{ab} \sim \eta^2 \kappa^8/J^4$. In comparison, the linear-order metric perturbation $h_{ab}^{(8)}$ from the quartic-Riemann operator is generated schematically via $\Box h_{ab}^{(8)} \sim \lambda {\cal R}^4$, so $h_{ab}^{(8)} \sim \lambda \kappa^6/J^3$ and analogously for $\tilde\lambda$. Hence, $(\delta h_{ab}^{(6)})/(h_{ab}^{(8)})\sim (\eta^2/\lambda)(\kappa^2/J)$, which we should be able to make arbitrarily small for sufficiently large black holes. Indeed, by the dispersion relation arguments of Ref.~\cite{Caron-Huot:2022ugt}, one has $\eta^2/\lambda \lesssim 1/(\kappa^2 \Lambda_{\rm UV}^2)$, for $\Lambda_{\rm UV}$ the energy scale of new physics, so as long as the size of the black hole is larger than the Compton wavelength of the ultraviolet states generating the higher-dimension operators (which must be the case to apply the EFT in the first place), we indeed find that $O(\eta^2)$ effects are negligible.}.
To study tidal force singularities, we are interested in the leading behavior of the Weyl tensor associated with the EFT-corrected metric. The effects of interest to this paper only become important near the horizon of a near-extremal black hole, since they arise from unusual  scaling in the near-horizon region when the horizon becomes arbitrarily far away (in spacelike directions). For this reason we can focus mainly on the near-horizon geometry. 
We proceed in two steps: i)~we EFT-correct the near-horizon extreme Kerr (NHEK) geometry~\cite{Bardeen:1999px}, and ii)~we then determine how metric perturbations fall off near the horizon by computing the so-called scaling dimensions $\gamma$.

We focus on extremal black hole solutions that are stationary and axisymmetric with respect to $k=\partial/\partial t$ and $m=\partial /\partial \phi$, respectively. Furthermore, we will impose the symmetry $(t,\phi)\to-(t,\phi)$. The most general ansatz compatible with the symmetries above can be written as 
\begin{equation}
\begin{aligned}
\mathrm{d}s^2 &= 2\,J\,\Omega^2\bigg[-\rho^2\mathrm{d}t^2+\frac{F_1}{\rho^2}\left(\mathrm{d}\rho+\rho\,F_2\,\mathrm{d}x\right)^2
\\
&\qquad\qquad\quad\qquad +\frac{\mathrm{d}x^2}{A}+B^2(\mathrm{d}\phi+\rho\,\omega\,\mathrm{d}t)^2\bigg],
\label{eq:general}
\end{aligned}
\end{equation}
where the factors of $\rho$ are adjusted so that $\rho=0$ is a Killing extremal horizon, $\phi\sim\phi+2\pi$, and $\Omega$, $\omega$, $A$, $B$, $F_1$, and $F_2$ are functions of $\rho$ and $x$, with $(\rho,x)\in \mathbb{R}^+\times [-1,1]$. To fix the gauge, we further impose $F_1=1$ and $F_2=0$. In the above, all coordinates are dimensionless, and in the case of the NHEK solution, $J$ is simply the Kerr angular momentum (in geometric units of length squared given by the size of the black hole).

\medskip

\noindent{\it a. EFT-corrected near-horizon geometries.---}In order to find the near-horizon geometry, we start by imposing ${\rm O}(2,1)\times {\rm U}(1)$ symmetry, in which case we take $\Omega=\Omega_{\rm NH}(x)$ and $B=B_{\rm NH}(x)$ in Eq.~(\ref{eq:general}) to be functions of $x$ only, as well as taking $A=A_{\rm NH}=(1-x^2)/\Gamma_{\rm NH}^2$, with $\Gamma_{\rm NH}$ and $\omega=\omega_{\rm NH}$ constants. The resulting line element reads
\begin{equation}
\begin{aligned}
\mathrm{d}s^2_{\rm NH} & =2\,J\,\Omega_{\rm NH}^2\bigg[-\rho^2\mathrm{d}t^2+\frac{\mathrm{d}\rho^2}{\rho^2}
+\frac{\Gamma_{\rm NH}^2\,\mathrm{d}x^2}{1-x^2}
\\
& \qquad\qquad\qquad\qquad +B_{\rm NH}^2(\mathrm{d}\phi+\rho\,\omega_{\rm NH}\,\mathrm{d}t)^2\Bigg].
\label{eq:near}
\end{aligned}
\end{equation}
The first two terms in brackets correspond to the metric on a two-dimensional unit-radius AdS$_2$, with $\rho=0$ being the location of the AdS$_2$ Poincar\'e horizon.

We expand all quantities appearing in Eq.~(\ref{eq:near}) as
\begin{equation}\label{eq:NHexpansion}
\hspace{-5mm}
\begin{aligned}
\Omega_{\rm NH} &{=} \Omega^{(0)}(x) \left[1{+}\eta\Omega^{(6)}(x) {+}\lambda\Omega^{(8)}(x) {+}\tilde{\lambda}\tilde{\Omega}^{(8)}(x)\right]
\\
B_{\rm NH}   &{=} B^{(0)}(x) \left[1{+}\eta B^{(6)}(x) {+}\lambda B^{(8)}(x) {+}\tilde{\lambda}\tilde{B}^{(8)}(x)\right]
\\
\Gamma_{\rm NH}   &{=} \Gamma^{(0)} \left[1{+}\eta\Gamma^{(6)}{+}\lambda \Gamma^{(8)}{+}\tilde{\lambda}\tilde{\Gamma}^{(8)}\right]
\\
 \omega_{\rm NH}  &{=} \omega^{(0)}\left[1{+}\eta\omega^{(6)}{+}\lambda\omega^{(8)}{+}\tilde{\lambda} \tilde{\omega}^{(8)}\right].
\end{aligned} \hspace{-5mm}
\end{equation}
For $\eta=\lambda=\tilde{\lambda}=0$ we recover the NHEK geometry, for which
\begin{equation}
\Omega^{(0)}(x) {=} \sqrt{\tfrac{1{+}x^2}{2}}, \; B^{(0)}(x)  {=} \tfrac{2\sqrt{1{-}x^2}}{1{+}x^2},\; \Gamma^{(0)}{=}\omega^{(0)}{=}1.
\end{equation}
Proceeding to the next order, we determine the corresponding EFT corrections. For instance, we find
\begin{equation}
\begin{aligned}
\Gamma^{(6)}&=-\frac{15\kappa^4}{32 \sqrt{2} J^2}, &\omega^{(6)}&=\frac{\kappa^4}{7 J^2},
\\
\Gamma^{(8)}&=-\frac{366435\,\kappa^6}{256 \sqrt{2} J^3}, &\omega^{(8)}&=\frac{(4864+1575 \pi)\,\kappa^6 }{20 J^3},
\\
\tilde{\Gamma}^{(8)}&=-\frac{368829\,\kappa^6}{64 \sqrt{2} J^3}, & \tilde{\omega}^{(8)}&=\frac{(4736+1575 \pi)\,\kappa^6 }{5 J^3},
\end{aligned}
\end{equation}
where some of these constants were chosen to ensure the absence of conical  singularities near the poles $x=\pm1$. Explicit expressions for $\Omega^{(I)}(x)$ and $B^{(I)}(x)$ can be found in the supplementary material (Sec.~\ref{sup:a}).

\medskip

\noindent{\it b. Deforming the EFT-corrected NHEK geometries.---} Having found the new near-horizon geometries, we now study how they respond to external tidal deformations. These deformations are generated by simply having the near-horizon geometry connected to an asymptotically flat region. We expect such deformations to become arbitrarily small near the extremal horizon, and as such we can use perturbation theory to study their properties.

We again start with Eq.~(\ref{eq:general}) with $F_1=1$ and $F_2=0$, but now $\Omega$, $B$, $A$ and $\omega$ can be functions of both $x$ and $\rho$. Since we are perturbing around an ${\rm O}(2,1)\times {\rm U}(1)$-symmetric solution and we are restricting our attention to stationary and axisymmetric perturbations, we decompose our perturbations in terms of harmonics on AdS$_2$ with dependence on $\rho$ only. These turn out to be power-law solutions of the form $\rho^\gamma$. We thus set 
\begin{equation}
\begin{aligned}
A(\rho,x) &=A_{\rm NH}(x)\left[1+\varepsilon\,\rho^\gamma\,Q_1(x)\right]
\\
B(\rho,x) &=B_{\rm NH}(x)\left[1+\varepsilon\,\rho^\gamma\,Q_2(x)\right]
\\
\Omega(\rho,x) &=\Omega_{\rm NH}(x)\left[1+\varepsilon\,\rho^\gamma\,Q_3(x)\right]
\\
\omega(\rho,x) &=\omega_{\rm NH}\left[1+\varepsilon\,\rho^\gamma\,Q_4(x)\right],
\end{aligned}
\end{equation}
with $\varepsilon$ being a bookkeeping parameter we take to be infinitesimally small. All the $Q_i$, with $i=1,2,3,4$, and $\gamma$ have an expansion in $\eta$, $\lambda$, and $\tilde{\lambda}$ of the form \footnote{Since at leading order field redefinitions change the metric by a smooth solution to Einstein's equations (which has integer $\gamma$), $\gamma^{(6)}, \gamma^{(8)}$, and $\tilde{\gamma}^{(8)}$ are field redefinition invariant.}
\begin{equation}
\hspace{-1mm}
\begin{aligned}
Q_i(x) &=Q^{(0)}_i(x){+}\eta Q^{(6)}_i(x){+}\lambda Q^{(8)}_i(x){+}\tilde{\lambda} \tilde{Q}^{(8)}_i(x)
\\
\gamma &=\gamma^{(0)}+\eta \gamma^{(6)}+\lambda \gamma^{(8)}+\tilde{\lambda} \tilde{\gamma}^{(8)}.
\end{aligned}
\hspace{-1mm}
\end{equation}
The main goal of this paper is to determine the corrections to the scaling dimensions $\gamma^{(6)}$, $\gamma^{(8)}$, and $\tilde{\gamma}^{(8)}$, since these control the deviation from the standard Kerr result. At each order, we find that $Q_2^{(I)}$ and $Q_3^{(I)}$ can be expressed in terms of  $Q_1^{(I)}$ and $Q_4^{(I)}$ and their first derivatives. We are thus left with two second-order differential equations for $Q_1^{(I)}$ and $Q_4^{(I)}$.

For $I=0$, i.e., deviations away from NHEK, the equations for $Q_1^{(0)}$ and $Q_4^{(0)}$ are second-order homogeneous equations of the Sturm-Liouville type where $\gamma^{(0)}$ appears as an eigenvalue. These can be solved analytically in terms of standard Legendre polynomials $P_{\ell}$ of degree $\ell$ (see Sec.~\ref{sup:b} of the supplementary material). We find two classes of solutions, labeled by $\gamma^{(0)}_{\pm}$, with critical exponents given by~\footnote{There is also an $\ell =1$  mode for $\gamma_+$ that is not pure gauge, but it does not contribute to tidal forces. It is analogous to the $\ell = 0 $ mode in spherically symmetric systems that changes $M$ or $Q$\label{ell1foot}}
\begin{equation}
\begin{aligned}
\gamma^{(0)}_+(\ell) &=\ell\, &\text{with}\quad \ell\in\mathbb{N}\geq 2,
\\
\gamma^{(0)}_-(\ell) &=\ell+1\, &\text{with}\quad \ell\in\mathbb{N}\geq 1.
\end{aligned}
\end{equation}
Since the scaling exponents are positive integers, the perturbation decays and the  horizon remains perfectly smooth. In other words, without higher-derivative corrections, the extremal Kerr horizon is unaffected by these deformations. It should be emphasized that these modes are already present for the asymptotically flat black holes.

For fixed values of $\ell$, there are two {\it distinct} values of $\gamma^{(0)}$, thus yielding a nondegenerate spectrum at fixed $\ell$ at zeroth order. We can thus proceed using standard perturbation theory. The resulting equations can again be solved analytically for each value of $\ell$, and after some algebra one finds
\begin{equation}
\begin{aligned}
\gamma^{(6)}_+(2)&=\gamma^{(6)}_-(1)=\frac{24\,\kappa^4}{7 J^2}
\\
\gamma^{(8)}_+(2)&=-\frac{21 (32+45 \pi)\kappa^6}{5 J^3} \\
\gamma^{(8)}_-(1)&=-\frac{9 (8576+3045 \pi )\kappa^6}{20 J^3}
\\
\tilde{\gamma}^{(8)}_+(2)&=-\frac{12 (736+315 \pi )\kappa^6}{5 J^3}
\\
\tilde{\gamma}^{(8)}_-(1)&=-\frac{189 (384+145 \pi )\kappa^6}{5 J^3}.
\end{aligned}
\end{equation}
At the moment we have no understanding of why $\gamma^{(6)}_-(1)=\gamma^{(6)}_+(2)$. Furthermore, $\gamma_-^{(I)}(\ell-1)\leq \gamma_+^{(I)}(\ell)$ for fixed values of $\ell$.

Even though these are small corrections, the scaling exponent  is no longer an integer, so when $\gamma < 2$ the curvature will diverge. The reader might worry that we have done this calculation using coordinates that are not regular at the extremal horizon. In the supplementary material in Sec.~\ref{sup:map}, we give an explicit map between the Kerr-like coordinates of Eq.~(\ref{eq:general}) and Bondi-Sachs coordinates, which are manifestly regular at $\rho=0$. It is in Bondi-Sachs coordinates that we compute the relevant components of the Weyl tensor and find diverging tidal force singularities at the future (and past) extremal event horizon so long as $\gamma<2$.
Since these modes were already present for the Kerr black hole, they are inevitably present if we simply assume that the full solution (beyond the near-horizon region) is also asymptotically flat. It should be noted that singular tidal forces at extremality can be linked to high curvature for nearly-extremal black holes. In particular, it was shown in Ref.~\cite{Horowitz:2022mly} that the curvature will diverge as  $T^{\gamma -2}$ in the limit $T \to 0$.

Since the metric is $C^1$ but not $C^2$, Einstein's equations remain well-defined in a distributional sense. However, since the event horizon is a Cauchy horizon for a constant-$t$ hypersurface, the evolution is not unique. We now turn to a situation where higher-curvature terms can have an even more dramatic effect.

\section{Extremal Reissner-Nordstr\"om}
The analogous calculation can be done for Reissner-Nordstr\"om black holes. The leading effective action now includes the following terms at fourth order in derivatives,
\begin{equation}\hspace{-1mm}
\begin{aligned}
    \mathcal{L}_4  &=  d_1 R^2 +  d_2 R_{a b}R^{a b} + d_3 R_{a b c d}R^{a b c d}
    \\
    &\;\;\; +\kappa^2 (d_4 R F^2 +    d_5 R^{a b} F^{2}_{ab} +  d_6 R^{a b c d} F_{a b} F_{c d})
    \\
  &\;\;\;  + \kappa^4 (d_7 (F^2)^2 +   d_8 F^2_{ab}F^{2ab}),
\end{aligned}\label{EFT:L4}\hspace{-1mm}
\end{equation}
where all $d_i$ are dimensionless and where we write $F^2 \equiv F_{ab}F^{ab}$ and $F^2_{ab}\equiv F^{\phantom{b}}_{ac}F_b^{\phantom{b}c}$.  As before, it suffices to focus on the near-horizon geometry, which is the Bertotti-Robinson solution. Since this is a well-defined and highly symmetrical solution in any spacetime dimension $D$, we will not restrict ourselves to $D\,{=}\,4$~\footnote{The results presented in this section were checked up to $D=11$ using the {\it xAct} suite of packages~\cite{xAct}.}. In the same way as in the previous section, we start by finding an EFT-corrected Bertotti-Robinson solution, keeping ${\rm O}(2,1) \times {\rm SO}(D\,{-}\,1)$ symmetry. The most general solution is still ${\rm AdS}_2 \times S^{D-2}$, with shifted radii of curvature $L_{{\rm AdS}_2}$ and $r_+$ depending on $Q$ and $d_i$,
\begin{equation}
\begin{aligned}
    {\rm d}s^2_{BR} &= L_{{\rm AdS}_2}^2 \left( -\rho^2 {\rm d}t^2 + \frac{ {\rm d}\rho^2}{\rho^2}\right) + r_+^2  {\rm d}\Omega_{D-2}^2 \\
    F &= \frac{Q L_{{\rm AdS}_2}}{r_+^2}  {\rm d}t \wedge  {\rm d}r,
    \end{aligned}
\end{equation}

Next, we consider the (EFT-corrected) linearized equation for static, nonspherical  perturbations, which could be sourced by distant matter. We may use the large symmetry at hand to decompose the perturbations into ${\rm AdS}_2$ harmonics and spherical harmonics. The former translates into simple power-law behavior $\delta g \sim \rho^\gamma$. The scalar-derived perturbations are given by:
\begin{equation}
\begin{aligned}
    \delta g_{IJ} &= s\,\mathbb{S}_{\ell}\,\rho^\gamma g_{IJ}  \\
    \delta g_{AB} &= \rho^\gamma \left( h_L\,\mathbb{S}_{\ell}\,g_{AB} + h_T\,\mathbb{S}_{AB}^\ell \right)  \\
    \delta F &= {\rm d}(q\,\mathbb{S}_\ell\,\rho^{\gamma+1}) \wedge  {\rm d}t,
\end{aligned}
\end{equation}
where indices $I,J$ and $A,B$ correspond to ${\rm AdS}_2$ and $S^{D-2}$, respectively, $\mathbb{S}_\ell$ is the ${\ell}$th  spherical harmonic, $S_{AB}^{\ell}$ is the traceless part of its second derivative, and $s, h_L, h_T, q$ are constants. The linearized equations then become an algebraic system for the coefficients that has solutions only for specific values of the exponent $\gamma$. The scaling dimension (for a given $\ell$) may take four values. Two of them are always negative and can be removed by boundary conditions at the horizon. For the smaller of the remaining two, we find
\begin{equation}
    \gamma(\ell) = \frac{\ell}{D{-}3} -1-d_0 \frac{8\kappa^2(D{-}4)\ell(\ell{+}D{-}3)}{(D{-}2)(D{-}3)^2(2\ell{-}D{+}3)r_+^2},\label{eq:gammaEM}
\end{equation}
where
\begin{equation}\hspace{-2mm}
\begin{aligned}
    d_0 &= \tfrac{(D-3) (D-4)^2}{4}  d_1+ \tfrac{(D-3) (2 D^2-11 D+16)}{4} d_2 \\&\;\; +\tfrac{(2 D^3-16 D^2+45 D-44)}{2} d_3 +\tfrac{(D-3) (D-2) (D-4)}{2}  d_4\\&\;\; +\tfrac{(D-2)(D-3)^2 }{2} (d_5 {+} d_6)+\tfrac{(D-2)^2 (D-3)}{2} (2d_7{+}d_8)
    \label{eq:d0}
\end{aligned}\hspace{-2mm}
\end{equation}
is a combination of Wilson coefficients invariant under arbitrary field redefinitions $g_{ab} \rightarrow g_{ab} + r_1 R_{ab} + r_2 R g_{ab} +  r_3 F^2_{ab} + r_4 F^2 g_{ab}$~\cite{Cheung:2018cwt}. We see that the EFT correction in Eq.~\eqref{eq:gammaEM} vanishes in four dimensions and, if $d_0 > 0$ as predicted by Ref.~\cite{Cheung:2018cwt}, becomes negative in higher dimensions. In particular, the $\ell=2$ mode in $D=5$ leads to a negative exponent, so the perturbation blows up at the horizon. This is a sign of the breakdown of perturbation theory, and the near-horizon geometry will no longer be ${\rm AdS}_2\times S^3$. This is analogous to what happens to nonspherical extremal black holes in ${\rm AdS}_5$~\cite{Horowitz:2022leb}.  As in that case, perturbations of the new horizon will likely still have noninteger scaling dimensions $0< \gamma < 2$, so there will still be curvature singularities.

\section{Discussion}
The fate of the black hole horizon depends sensitively on the signs of the coefficients of the higher-dimension operators.
From our calculation of the scaling dimensions, we see that extremal Kerr black hole horizons develop singularities if $\eta$ is negative, or if $\lambda$ or $\tilde\lambda$ are positive. 
 Further, if $d_0>0$, then the extremal RN black hole in five dimensions develops a singular horizon~\footnote{The significance of the negative shift in $\gamma$, for $d_0>0$, for extremal charged black holes in $D\geq 6$ is less clear, since generic nonspherical perturbations blow up on the horizon already in the pure Einstein-Maxwell theory.}.
The signs of the various Wilson coefficients in the EFT are therefore of paramount importance in determining the nature of the horizon.

Fortuitously, there exists an infrared consistency program of bounding the coefficients of EFTs from first principles using the tools of analytic dispersion relations in quantum field theory, in a manner agnostic of the details of the ultraviolet theory~\cite{Adams:2006sv}.
Using either analyticity and unitarity of scattering amplitudes~\cite{Bellazzini:2015cra} or causality of graviton propagation~\cite{Gruzinov:2006ie}, one can prove that, in any consistent theory of quantum gravity, the coefficients of the quartic Riemann operators $\lambda$ and $\tilde\lambda$ must be positive.
This is borne out in string theory~\cite{Bellazzini:2015cra}, where calculations of the low-energy EFT yield $\lambda = \alpha'^3[13+\zeta(3)]/256\kappa^6$ and $\tilde\lambda = \alpha'^3[1+\zeta(3)]/1024\kappa^6$ in the bosonic case~\cite{Jack:1988sw,Jack:1989vp},
$\lambda = 4\tilde\lambda = \alpha'^3 \zeta(3)/256\kappa^6$ in the type~II case~\cite{Gross1986,Gross:1986mw,Kikuchi:1986rk}, and $\lambda = 4\tilde\lambda = \alpha'^3 [1+2\zeta(3)]/512\kappa^6$ in the heterotic~\cite{Gross:1986mw,Kikuchi:1986rk} (equivalently, type~I~\cite{Tseytlin:1995bi}) case.

However, the Riemann-cubed operator is not amenable to such a general theory-independent bound on the basis of extant dispersion relations.
While vanishing on general grounds in any supersymmetric theory~\cite{Camanho:2014apa}, the $\eta$ coefficient receives finite contributions at one loop~\cite{Goon:2016mil} from massive states in the theory~\footnote{At two loops, $\eta$ is generated even in a purely gravitational theory, with a beta function that leads to positive $\eta$ in the asymptotic infrared~\cite{Bern:2015xsa,Bern:2017puu}, though this effect is too small to be physically relevant, even for astrophysical black holes.},
\begin{equation}
\eta = \frac{1}{15120(4\pi)^2\kappa^2} \sum\left(\frac{1}{m_{{\rm s}}^2}-\frac{4}{m_{{\rm f}}^2}+\frac{3}{m_{{\rm v}}^2}\right), 
\end{equation}
where $m_{{\rm s}}$, $m_{{\rm f}}$, and $m_{{\rm v}}$ label the masses of the heavy real scalars, Dirac fermions, and vectors in the theory. As expected, this combination vanishes for supersymmetric theories.
In the standard model, where the lightest massive states are fermions, we therefore will have $\eta < 0$, leading to singular horizons for rapidly-spinning Kerr black holes.
Conversely, if the near-horizon region of a rapidly spinning black hole were observed to be regular, this would constitute evidence for new light hidden-sector bosons. 

Further, the signs of Wilson coefficients in the Einstein-Maxwell EFT are of great importance for the Weak Gravity Conjecture (WGC)~\cite{Arkani-Hamed:2006emk,Kats:2006xp,Cheung:2018cwt,Cheung:2019cwi,Arkani-Hamed:2021ajd} and black hole thermodynamics~\cite{Cheung:2018cwt,Cheung:2019cwi,Arkani-Hamed:2021ajd,Reall:2019sah,Goon:2019faz}.
The WGC is the statement that, when an Abelian gauge theory is consistently coupled to quantum gravity, there should always be a state in the spectrum for which $Q/M>1$ in Planck units, with $1$ corresponding to the charge-to-mass ratio of an extremal black hole~\cite{Arkani-Hamed:2006emk}.
In the presence of higher-dimension operators in the EFT, the black hole solutions are deformed, and thus the maximum allowed charge-to-mass ratio of black holes is also shifted.
For electric black holes, the particular combination $d_0$ defined in Eq.~\eqref{eq:d0} dictates this shift, with $\Delta(Q/M) \propto d_0$~\cite{Cheung:2018cwt}, so that $d_0 > 0$ allows the black holes themselves to satisfy the WGC.
In this case, there are expected to be nonperturbative decay processes, i.e., a black hole version of Schwinger pair production, by which large extremal black holes decay to smaller ones with $Q/M$ slightly $>1$.
Interestingly, the shift in the black hole entropy and on-shell action are also $\propto d_0$~\cite{Cheung:2018cwt,Arkani-Hamed:2021ajd,Goon:2019faz}, and it was argued via black hole thermodynamics in Ref.~\cite{Cheung:2018cwt} that $d_0>0$ in tree-level completions.
It is noteworthy that, for precisely this same situation $d_0>0$, we also find an instability toward singular extremal charged black holes in $D=5$, along with a negative shift in the associated scaling dimension for all $D\geq 5$.
We leave the question of whether this effect is somehow connected with the WGC-mandated black hole decay to future work.

When the curvature diverges on the horizon of an extremal black hole, a near-extremal black hole has curvature that diverges as one approaches extremality~\cite{Horowitz:2022mly}. 
Since we observe rapidly spinning black holes in nature (see, e.g., Refs.~\cite{Gou:2011nq, Risaliti:2013cga,Zackay:2019tzo}), one might wonder if the effect described here could be observable. 
That is possible, but detecting tidal forces much larger than the ambient curvature  
would require black holes much closer to extremality than have been observed to date. 
However, we take our results as a proof of principle that black hole horizons can serve to amplify the effects of higher-derivative terms in the action, leading to a surprising breakdown of the EFT in the near-horizon region, and we leave the broader question of possibly realizing this effect astrophysically to future work.

\medskip

\noindent {\it Note added:} Recently, Cano and David~\cite{Cano:2024bhh} have generalized our results, showing that the parity-odd quartic term ${\cal C}\tilde {\cal C}$ does indeed contribute to the scaling exponents parameterizing the diverging tidal forces on the extremal Kerr horizon, while the parity-odd Riemann-cubed term does not.
We set the Wilson coefficients of these operators to zero throughout.

\medskip

\noindent {\it Acknowledgments:} We thank Harvey Reall for early collaboration, as well as Justin Berman and Cliff Cheung for discussions. The work of G.H. was supported in part by NSF Grant PHY-2107939. M.K. was partially supported by Polish budgetary funds for science in 2018-2022 as a research project under the program ``Diamentowy Grant.''
G.N.R. is supported at the Kavli Institute for Theoretical Physics by the Simons Foundation (Grant No.~216179) and the National Science Foundation (Grant No.~NSF PHY-1748958) and at the University of California, Santa Barbara by the Fundamental Physics Fellowship. J.E.S. has been partially supported by STFC consolidated grant ST/T000694/1.

\bibliographystyle{utphys-modified}
\bibliography{EFT}
\clearpage
\widetext
\input{final_sup}
\end{document}

%% file: final_sup.tex
\section{Supplementary Material}
\subsection{\label{sup:a}EFT-corrected near-horizon geometries}
Let $C^{(6)}$, $C^{(8)}$, and $\tilde{C}^{(8)}$ be integration constants, and define $K(x)\equiv\arcsin\left(\frac{\sqrt{2} x}{\sqrt{1+x^2}}\right)-\arcsin x$.  
Below, we give the explicit form of the terms that appear in Eq.~\eqref{eq:NHexpansion}.

\subsubsection*{$I=6$}

For the six-derivative corrections, we find 
\begin{multline}
B^{(6)}(x)=\frac{\kappa^4}{J^2}\Bigg[\frac{2656-42885 x^2+45895 x^4-8130 x^6-1218 x^8+ 183 x^{10}+139 x^{12}}{224(1+x^2)^6}
-\frac{15 \sqrt{2} x (3-x^2) }{64(1+x^2)\sqrt{1-x^2}} K(x)\Bigg]
\end{multline}
and 
\begin{equation}
\Omega^{(6)}(x)=\frac{\kappa^4}{J^2} \Bigg[C^{(6)}-\frac{3285-55449 x^2+54210 x^4-7058 x^6-1527 x^8-309 x^{10}}{224 (1+x^2)^6}
+\frac{15 x \sqrt{2} \sqrt{1-x^2}}{64 (1+x^2)}K(x)\Bigg].
\end{equation}

\subsubsection*{$I=8$}
For the eight-derivative corrections, there are two families of solutions: 
\begin{equation}
\begin{aligned}
B^{(8)}(x) &=\frac{\kappa^6}{J^3} \Bigg[\frac{832989}{1280}-\frac{315 \pi }{4}
-\frac{407005+32887800 x^2+38302380 x^4+227158536 x^6}{1280 (1+x^2)^9}
\\
&\qquad\qquad -\frac{244951182 x^8+207667400 x^{10}+108083820 x^{12}+31954360 x^{14}+4114685 x^{16}}{1280 (1+x^2)^9}
\\
&\qquad\qquad +\frac{630 x}{(1+x^2)}\arctan x-\frac{366435 x (3-x^2)}{256 \sqrt{2} \sqrt{1-x^2} (1+x^2)} K(x)\Bigg],
\end{aligned}
\end{equation}
\begin{equation}
\begin{aligned}
\Omega^{(8)}(x) &=\frac{\kappa^6}{J^3} \Bigg[C^{(8)}+\frac{783837+16684758 x^2+33602022 x^4+119986542 x^6}{1280 (1+x^2)^9}
\\
&\qquad\qquad +\frac{27639936 x^8+23049562 x^{10}+11880370 x^{12}+3484978 x^{14}+445863 x^{16}}{256 (1+x^2)^9}
\\&\qquad\qquad -\frac{315 x}{1+x^2}\arctan x+\frac{366435 x \sqrt{1-x^2}}{256
   \sqrt{2} (1+x^2)}K(x)\Bigg],
\end{aligned}
\end{equation}
and 
\begin{equation}
\begin{aligned}
\tilde{B}^{(8)}(x)&=\frac{\kappa^6}{J^3} \Bigg[\frac{846339}{320}-315 \pi-\frac{1149443+5618952 x^2+136013268 x^4+154320120 x^6+254641842 x^8}{320 (1+x^2)^9}
\\
&\qquad\qquad -\frac{208733752 x^{10}+108674580 x^{12}+32136008 x^{14}+4138723
   x^{16}}{320 (1+x^2)^9}
   \\&\qquad\qquad +\frac{2520 x}{1+x^2}\arctan x-\frac{368829 x (3-x^2)}{64
   \sqrt{2} \sqrt{1-x^2} (1+x^2)}K(x)\Bigg],
\end{aligned}
\end{equation}
\begin{equation}
\begin{aligned}
\tilde{\Omega}^{(8)}(x)&=\frac{\kappa^6}{J^3} \Bigg[\tilde{C}^{(8)}+\frac{1018371+7724394 x^2+67516506 x^4+96062418 x^6+141833088 x^8}{320 (1+x^2)^9}
\\
&\qquad\qquad +\frac{115923454 x^{10}+59757382 x^{12}+17530822 x^{14}+2243037 x^{16}}{320
   (1+x^2)^9}
   \\&\qquad\qquad -\frac{1260 x}{1+x^2}\arctan x+\frac{368829 x \sqrt{1-x^2}}{64 \sqrt{2}
   (1+x^2)}K(x)\Bigg].
\end{aligned}
\end{equation}

\subsection{\label{sup:b}Approach to the  near-horizon geometry}
We find two families of modes for stationary and axisymmetric deformations of the near-horizon limit of the extremal Kerr black hole.

For the $+$ family, we find 
\begin{equation} 
\begin{aligned}
\gamma_+^{(0)}(\ell) &=\ell
\\
Q^{(0)}_{1\;\;+}(x) &=P^\prime_\ell(x)
\\
Q^{(0)}_{2\;\;+}(x) &=-\frac{1}{2 (1+x^2)}\Bigg[2\ell(\ell+1)\,x\,P_{\ell }(x)+(1-3x^2) P_{\ell }'(x)\Bigg]
\\
Q^{(0)}_{3\;\;+}(x) &=\frac{1}{2 (1+x^2)}\Bigg[\ell(\ell+1)\,x\,P_{\ell }(x)+(1-x^2) P_{\ell }'(x)\Bigg]
\\
Q^{(0)}_{4\;\;+}(x) &=\frac{1}{2} \Bigg[\ell\,x\,P_{\ell }(x)+\frac{(1-x^2) (\ell^2+\ell+2)}{2 (\ell+1 )} P_{\ell }'(x)\Bigg],
\end{aligned}
\end{equation}
where ${}^\prime$ denotes differentiation with respect to $x$. Modes with $\ell=0$ vanish, while modes with $\ell=1$ are special~\hyperref[ell1foot]{[14]},   so that in the $+$ family we take $\ell\geq2$.

For the $-$ family, we find
\begin{equation}
\begin{aligned}
\gamma_-^{(0)}(\ell) &=\ell+1
\\
Q^{(0)}_{1\;\;-}(x)&=0
\\
Q^{(0)}_{2\;\;-}(x) &=\frac{1-x^2}{1+x^2} \Bigg\{\ell  (\ell +1)\,x\,P_{\ell }(x)-\left[2+(1-x^2) \ell \right] P_{\ell }'(x)\Bigg\}
\\
Q^{(0)}_{3\;\;-}(x) &=-\frac{1}{2}\frac{1-x^2}{1+x^2} \Bigg\{\ell  (\ell +1)\,x\,P_{\ell }(x)-\left[2+(1-x^2) \ell \right] P_{\ell }'(x)\Bigg\}
\\
Q^{(0)}_{4\;\;-}(x) &=\ell  (\ell +1) \,x\,P_{\ell }(x)+(1+x^2 \ell)\,P_{\ell }'(x),
\end{aligned}
\end{equation}
where modes with $\ell=0$ vanish, so that for the $-$ family we take $\ell\geq1$.

\subsubsection*{Two examples of EFT-corrected deformations}
 We normalize perturbations so that, for each value of $\ell$,
\begin{equation}
\int_{-1}^1 Q_{1\;\;+}(x)^2\mathrm{d}x=\ell(\ell+1)\quad\text{and}\quad \int_{-1}^1\frac{(1-x^2)^2}{x^2} Q_{4\;\;-}^\prime(x)^2\mathrm{d}x=\frac{2 \ell  (\ell +1)^3 (\ell +2) (\ell +3)}{2 \ell +3}.
\end{equation}

We give examples of EFT corrections in the $+$ family, with $\ell=2$. For the six-derivative correction, we find
\begin{equation}
\begin{aligned}
Q_{1\;\;+}^{(6)}(x)=\frac{\kappa^4}{J^2} \frac{3x}{224}\Bigg\{35
   \sqrt{2}-1675-\frac{36372480}{(1+x^2)^7}+\frac{72826880}{(1+x^2)^6}-\frac{50065408}{(1+x^2)^5}+\frac{13659904}{(1+x^2)^4}-\frac{1148992}{(1+x^2)^3}
   \\-\frac{13
   6}{(1+x^2)^2}-\frac{374}{1+x^2}
   +384 \log\left[\frac{1}{2}(1+x^2)\right]+\frac{105 \sqrt{1-x^2} }{\sqrt{2}}\frac{K(x)}{x}\Bigg\},
\end{aligned}
\end{equation}
while for one of the eight-derivative corrections, we have
\begin{equation}
\begin{aligned}
Q_{1\;\;+}^{(8)}(x)=\frac{\kappa^6}{J^3} \frac{9x}{20}\Bigg\{\frac{203230547-2721885065 x^2+10948541740 x^4-13485253140 x^6+8456364570 x^8}{480 (1+x^2)^{10}}
\\
-\frac{729506558 x^{10}-621146940 x^{12}-217212700 x^{14}-50262315
   x^{16}-5216895 x^{18}}{480 (1+x^2)^{10}}
   \\
   +\frac{420 \pi }{1-x^2}-\frac{4462619}{320}+\frac{203575}{32 \sqrt{2}}+2520 C -2985 \pi  +\frac{610725 \sqrt{1-x^2}}{64 \sqrt{2}}\frac{K(x)}{x} \\ +\frac{1680 (2-3 x^2)}{(1-x^2)}\frac{\arctan x}{x}
   -5040 \arctan x\;{\rm arctanh} x -448
   \log \left(\frac{1+x^2}{2}\right) \\ -630 \pi  \log \left(\frac{1-x^2}{4}\right)
   -2520\;D\left[\left(\frac{1+i}{2}\right) (1-x)\right]-2520\;
   D\left[\left(\frac{1+i}{2}\right) (1+x)\right]\Bigg\},
\end{aligned}
\end{equation}
where $D(z)$ is the Bloch-Wigner dilogarithm function defined as
\begin{equation}
D(z)={\rm Im}({\rm Li}_2(z))+ {\rm arg}(1-z)\,\log|z|,
\end{equation}
${\rm Li}_2(z)$ is the dilogarithm function, and $-\pi\leq \arg z< \pi$. Note that $D(z)$ is a real analytic function on $\mathbb{C}$ except at the two points $z=0$ and $z=1$, where it is continuous but not differentiable (with singularities of the form $|z| \log |z|$ and $|1-z|\log|1-z|$, respectively). Finally, $C = \sum_{n=0}^\infty (-1)^n/(2n+1)^2 \approx 0.915966$ is Catalan's constant.
\subsection{\label{sup:map}Map between perturbations in Kerr-like and Bondi-Sachs coordinates}
One might worry that the coordinates we  used to determine the behavior of near-horizon deformations away from the EFT-corrected near-horizon geometries are not regular at the event horizon. Indeed, this is the case even for the Kerr-like coordinates used in Eq.~(\ref{eq:near}). In this section, we show that a simple map exists between the Kerr-like coordinates of Eq.~(\ref{eq:general}) (with $F_1=1$ and $F_2=0$) and standard Bondi-Sachs coordinates. This map was worked out as an expansion in $\varepsilon$.

Recall that Bondi-Sachs coordinates take the following general form:
\begin{equation}
\mathrm{d}s^2=(-V\mathrm{d}v^2+2\,\mathrm{d}v\,\mathrm{d}\rho)e^{2\beta}+e^{2\chi}h_{\dot{p}\dot{q}}(\mathrm{d}y^{\dot{p}}+U^{\dot{p}}\mathrm{d}v)(\mathrm{d}y^{\dot{q}}+U^{\dot{q}}\mathrm{d}v)
\label{eq:bondi}
\end{equation}
with $\dot{p},\,\dot{q}=1,2$.
Consider the following coordinate transformation:
\begin{equation}
\begin{aligned}
t &=v+\frac{1}{\rho}
\\
\phi &= \varphi+\log \rho+\lambda(\rho,x).
\end{aligned}
\end{equation}
One finds that Eq.~(\ref{eq:general}) takes the same form as Eq.~(\ref{eq:bondi}) with $V=\rho^2$, $y^{\dot{p}}=\{x,\varphi\}$, $e^{2\beta}=e^{2\chi}=2\,J\,\Omega ^2$, and
\begin{equation}
\begin{aligned} 
h_{\dot{p}\dot{q}} &=\left[
\begin{array}{cc}
\displaystyle \frac{1}{A}+B^2 \left(\frac{\partial \lambda}{\partial x}\right)^2 & \displaystyle -B^2\left(\frac{\partial \lambda}{\partial x}\right)
\\
\\
 \displaystyle -B^2\left(\frac{\partial \lambda}{\partial x}\right) &\displaystyle B^2
\end{array}
\right] \\
U^{\dot{p}} &=\left[\begin{array}{cc}
\displaystyle0 & \displaystyle\rho\,\omega
\end{array}\right],
\end{aligned}
\end{equation}
so long as
\begin{equation}
\frac{\partial \lambda}{\partial \rho}=\frac{1-\omega}{\rho}.
\end{equation}
The last relation is a first-order equation in $\rho$ and can be readily solved for any $\omega$.

Since $V=\rho^2$, in Bondi-Sachs coordinates, the future extremal event horizon is the null hypersurface $\rho=0$. So long as $\gamma>0$, these coordinates are regular at the horizon and can be used to extend the deformations of the EFT near-horizon geometries to a new region with $\rho<0$. Since for extremal black holes the (past and future) event horizons are also Cauchy surfaces for initial data on constant-$t$ hypersurfaces, the extension beyond $\rho=0$ into the new region with $\rho<0$ is not {unique}.

It is now a simple exercise to show that, to linear order in $\varepsilon$, the Weyl tensor satisfies
\begin{equation}
C_{\rho \dot{p} \rho \dot{q}}= J\,(1-\gamma)\,\rho^{\gamma-2}\,\varepsilon\,\left[
\begin{array}{cc}
\displaystyle -\frac{\gamma\,\Gamma_{\rm NH}^2(Q_1+2 Q_2)\Omega_{\rm NH}^2}{2(1-x^2)} & \displaystyle B_{\rm NH}^2 \Omega_{\rm NH}^2 Q_4^\prime
\\
\\
\displaystyle B_{\rm NH}^2 \Omega_{\rm NH}^2 Q_4^\prime & \displaystyle -\frac{\gamma\,B_{\rm NH}^2(Q_1+2 Q_2)\Omega_{\rm NH}^2}{2} 
\end{array}\right].
\end{equation}
For Kerr, the smallest scaling exponent that contributes to the above quantity has $\gamma=2$, rendering $C_{\rho \dot{p} \rho \dot{q}}$ finite. But when higher-derivative corrections are included, $\gamma < 2$ and the curvature diverges, resulting in infinite tidal forces.

%% file: final.bbl
\providecommand{\href}[2]{#2}\begingroup\raggedright\begin{thebibliography}{10}

\bibitem{Note1}
This effect is different from the dynamical instability of extremal black holes
  discussed in Refs.~\cite {Aretakis:2012ei,Gajic:2023uwh}.

\bibitem{Horowitz:2022mly}
G.~T. Horowitz, M.~Kolanowski, and J.~E. Santos, ``{Almost all extremal black
  holes in AdS are singular},''
  \href{http://dx.doi.org/10.1007/JHEP01(2023)162}{{\em JHEP} {\bfseries 01}
  (2023) 162}, \href{http://arxiv.org/abs/2210.02473}{{\ttfamily
  arXiv:2210.02473 [hep-th]}}.

\bibitem{Chen:2018jed}
B.~Chen and L.~C. Stein, ``{Deformation of extremal black holes from stringy
  interactions},'' \href{http://dx.doi.org/10.1103/PhysRevD.97.084012}{{\em
  Phys. Rev. D} {\bfseries 97} (2018) 084012},
  \href{http://arxiv.org/abs/1802.02159}{{\ttfamily arXiv:1802.02159 [gr-qc]}}.

\bibitem{Fulling}
S.~Fulling, R.~C. King, B.~Wybourne, and C.~Cummins, ``{Normal forms for tensor
  polynomials. 1: The Riemann tensor},''
\href{http://dx.doi.org/10.1088/0264-9381/9/5/003}{{\em Class.Quant.Grav.}
  {\bfseries 9} (1992) 1151}.

\bibitem{Note2}
We are only counting parity-even terms, since one can show that parity-odd
  terms do not affect our result.

\bibitem{Endlich:2017tqa}
S.~Endlich, V.~Gorbenko, J.~Huang, and L.~Senatore, ``{An effective formalism
  for testing extensions to General Relativity with gravitational waves},''
  \href{http://dx.doi.org/10.1007/JHEP09(2017)122}{{\em JHEP} {\bfseries 09}
  (2017) 122}, \href{http://arxiv.org/abs/1704.01590}{{\ttfamily
  arXiv:1704.01590 [gr-qc]}}.

\bibitem{Note3}
We work with the sign conventions $\epsilon _{0123}=\protect \sqrt {-g}$,
  $R_{ab} = R^c_{\protect \tmspace +\thickmuskip {.2777em}\protect \tmspace
  +\thickmuskip {.2777em}acb}$, $R^a_{\protect \tmspace +\thickmuskip
  {.2777em}\protect \tmspace +\thickmuskip {.2777em}bcd} = \partial _c \Gamma
  ^a_{bd} - \partial _d \Gamma ^a_{bc} + \Gamma ^a_{ce}\Gamma ^e_{bd} - \Gamma
  ^a_{de}\Gamma ^e_{bc}$, $\Gamma ^a_{bc} = g^{ad}(\partial _c g_{db} +
  \partial _b g_{dc} - \partial _d g_{bc})/2$, and mostly-plus metric
  signature.

\bibitem{Cardoso:2018ptl}
V.~Cardoso, M.~Kimura, A.~Maselli, and L.~Senatore, ``{Black Holes in an
  Effective Field Theory Extension of General Relativity},''
  \href{http://dx.doi.org/10.1103/PhysRevLett.121.251105}{{\em Phys. Rev.
  Lett.} {\bfseries 121} (2018) 251105},
  \href{http://arxiv.org/abs/1808.08962}{{\ttfamily arXiv:1808.08962 [gr-qc]}}.

\bibitem{Cano:2019ore}
P.~A. Cano and A.~Ruip\'erez, ``{Leading higher-derivative corrections to Kerr
  geometry},'' \href{http://dx.doi.org/10.1007/JHEP05(2019)189}{{\em JHEP}
  {\bfseries 05} (2019) 189}, \href{http://arxiv.org/abs/1901.01315}{{\ttfamily
  arXiv:1901.01315 [gr-qc]}}.
  \href{https://doi.org/10.1007/JHEP03(2020)187}{[Erratum: {\it JHEP} {\bf 03}
  (2020) 187]}.

\bibitem{Reall:2019sah}
H.~S. Reall and J.~E. Santos, ``{Higher derivative corrections to Kerr black
  hole thermodynamics},'' \href{http://dx.doi.org/10.1007/JHEP04(2019)021}{{\em
  JHEP} {\bfseries 04} (2019) 021},
  \href{http://arxiv.org/abs/1901.11535}{{\ttfamily arXiv:1901.11535
  [hep-th]}}.

\bibitem{Note4}
We note that the second-order perturbation $\delta h^{(6)}_{ab}$ from
  ${\protect \cal R}^3$ at $O(\eta ^2)$ is smaller than the linear-order
  quartic-Riemann perturbation $h_{ab}^{(8)}$ via the following reasoning. The
  linear perturbation from the cubic-Riemann term, by Eqs.~\protect \eqref
  {eq:EOMpre} and \protect \eqref {eq:EOM}, is given schematically by $\Box
  h^{(6)}_{ab} \sim \eta {\protect \cal R}^3$, so $h^{(6)}_{ab} \sim \eta
  \kappa ^4/J^2$. Then the $O(\eta ^2)$ backreaction effect is generated via
  the equation of motion as $\Box \delta h^{(6)}_{ab} \sim (\nabla
  h_{ab}^{(6)})^2 + \eta {\protect \cal R}^2 \delta {\protect \cal R} \sim \eta
  ^2 \kappa ^8/J^5$, so $\delta h^{(6)}_{ab} \sim \eta ^2 \kappa ^8/J^4$. In
  comparison, the linear-order metric perturbation $h_{ab}^{(8)}$ from the
  quartic-Riemann operator is generated schematically via $\Box h_{ab}^{(8)}
  \sim \lambda {\protect \cal R}^4$, so $h_{ab}^{(8)} \sim \lambda \kappa
  ^6/J^3$ and analogously for $\protect \tilde \lambda $. Hence, $(\delta
  h_{ab}^{(6)})/(h_{ab}^{(8)})\sim (\eta ^2/\lambda )(\kappa ^2/J)$, which we
  should be able to make arbitrarily small for sufficiently large black holes.
  Indeed, by the dispersion relation arguments of Ref.~\cite
  {Caron-Huot:2022ugt}, one has $\eta ^2/\lambda \lesssim 1/(\kappa ^2 \Lambda
  _{\protect \rm UV}^2)$, for $\Lambda _{\protect \rm UV}$ the energy scale of
  new physics, so as long as the size of the black hole is larger than the
  Compton wavelength of the ultraviolet states generating the higher-dimension
  operators (which must be the case to apply the EFT in the first place), we
  indeed find that $O(\eta ^2)$ effects are negligible.

\bibitem{Bardeen:1999px}
J.~M. Bardeen and G.~T. Horowitz, ``{Extreme Kerr throat geometry: A vacuum
  analog of ${\rm AdS}(2)\times S^2$},''
  \href{http://dx.doi.org/10.1103/PhysRevD.60.104030}{{\em Phys. Rev. D}
  {\bfseries 60} (1999) 104030},
  \href{http://arxiv.org/abs/hep-th/9905099}{{\ttfamily arXiv:hep-th/9905099}}.

\bibitem{Note5}
Since at leading order field redefinitions change the metric by a smooth
  solution to Einstein's equations (which has integer $\gamma $), $\gamma
  ^{(6)}, \gamma ^{(8)}$, and $\protect \tilde {\gamma }^{(8)}$ are field
  redefinition invariant.

\bibitem{Note6}
There is also an $\ell =1$ mode for $\gamma _+$ that is not pure gauge, but it
  does not contribute to tidal forces. It is analogous to the $\ell = 0 $ mode
  in spherically symmetric systems that changes $M$ or $Q$\label {ell1foot}.

\bibitem{Note7}
The results presented in this section were checked up to $D=11$ using the
  {\protect \it xAct} suite of packages~\cite {xAct}.

\bibitem{Cheung:2018cwt}
C.~Cheung, J.~Liu, and G.~N. Remmen, ``{Proof of the Weak Gravity Conjecture
  from Black Hole Entropy},''
  \href{http://dx.doi.org/10.1007/JHEP10(2018)004}{{\em JHEP} {\bfseries 10}
  (2018) 004}, \href{http://arxiv.org/abs/1801.08546}{{\ttfamily
  arXiv:1801.08546 [hep-th]}}.

\bibitem{Horowitz:2022leb}
G.~T. Horowitz, M.~Kolanowski, and J.~E. Santos, ``{A deformed IR: a new IR
  fixed point for four-dimensional holographic theories},''
  \href{http://dx.doi.org/10.1007/JHEP02(2023)152}{{\em JHEP} {\bfseries 02}
  (2023) 152}, \href{http://arxiv.org/abs/2211.01385}{{\ttfamily
  arXiv:2211.01385 [hep-th]}}.

\bibitem{Note8}
The significance of the negative shift in $\gamma $, for $d_0>0$, for extremal
  charged black holes in $D\geq 6$ is less clear, since generic nonspherical
  perturbations blow up on the horizon already in the pure Einstein-Maxwell
  theory.

\bibitem{Adams:2006sv}
A.~Adams, N.~Arkani-Hamed, S.~Dubovsky, A.~Nicolis, and R.~Rattazzi,
  ``{Causality, analyticity and an IR obstruction to UV completion},''
  \href{http://dx.doi.org/10.1088/1126-6708/2006/10/014}{{\em JHEP} {\bfseries
  10} (2006) 014}, \href{http://arxiv.org/abs/hep-th/0602178}{{\ttfamily
  arXiv:hep-th/0602178}}.

\bibitem{Bellazzini:2015cra}
B.~Bellazzini, C.~Cheung, and G.~N. Remmen, ``{Quantum Gravity Constraints from
  Unitarity and Analyticity},''
  \href{http://dx.doi.org/10.1103/PhysRevD.93.064076}{{\em Phys. Rev. D}
  {\bfseries 93} (2016) 064076},
  \href{http://arxiv.org/abs/1509.00851}{{\ttfamily arXiv:1509.00851
  [hep-th]}}.

\bibitem{Gruzinov:2006ie}
A.~Gruzinov and M.~Kleban, ``{Causality Constrains Higher Curvature Corrections
  to Gravity},'' \href{http://dx.doi.org/10.1088/0264-9381/24/13/N02}{{\em
  Class. Quant. Grav.} {\bfseries 24} (2007) 3521--3524},
  \href{http://arxiv.org/abs/hep-th/0612015}{{\ttfamily arXiv:hep-th/0612015}}.

\bibitem{Jack:1988sw}
I.~Jack, D.~R.~T. Jones, and N.~Mohammedi, ``{The four-loop metric
  $\beta$-function for the bosonic $\sigma$-model},''
  \href{http://dx.doi.org/10.1016/0370-2693(89)90031-2}{{\em Phys. Lett. B}
  {\bfseries 220} (1989) 171}.

\bibitem{Jack:1989vp}
I.~Jack, D.~R.~T. Jones, and N.~Mohammedi, ``{A four-loop calculation of the
  metric $\beta$-function for the bosonic $\sigma$-model and the string
  effective action},''
  \href{http://dx.doi.org/10.1016/0550-3213(89)90422-7}{{\em Nucl. Phys. B}
  {\bfseries 322} (1989) 431}.

\bibitem{Gross1986}
D.~J. Gross and E.~Witten, ``{Superstring modifications of Einstein's
  equations},''
  \href{http://dx.doi.org/https://doi.org/10.1016/0550-3213(86)90429-3}{{\em
  Nucl. Phys. B} {\bfseries 277} (1986) 1}.

\bibitem{Gross:1986mw}
D.~J. Gross and J.~H. Sloan, ``{The quartic effective action for the heterotic
  string},'' \href{http://dx.doi.org/10.1016/0550-3213(87)90465-2}{{\em Nucl.
  Phys. B} {\bfseries 291} (1987) 41}.

\bibitem{Kikuchi:1986rk}
Y.~Kikuchi, C.~Marzban, and Y.~J. Ng, ``{Heterotic string modifications of
  Einstein's and Yang-Mills' actions},''
  \href{http://dx.doi.org/10.1016/0370-2693(86)90924-X}{{\em Phys. Lett. B}
  {\bfseries 176} (1986) 57}.

\bibitem{Tseytlin:1995bi}
A.~A. Tseytlin, ``{Heterotic -- type I superstring duality and low-energy
  effective actions},''
  \href{http://dx.doi.org/10.1016/0550-3213(96)00080-6}{{\em Nucl. Phys. B}
  {\bfseries 467} (1996) 383},
  \href{http://arxiv.org/abs/hep-th/9512081}{{\ttfamily arXiv:hep-th/9512081}}.

\bibitem{Camanho:2014apa}
X.~O. Camanho, J.~D. Edelstein, J.~Maldacena, and A.~Zhiboedov, ``{Causality
  Constraints on Corrections to the Graviton Three-Point Coupling},''
  \href{http://dx.doi.org/10.1007/JHEP02(2016)020}{{\em JHEP} {\bfseries 02}
  (2016) 020}, \href{http://arxiv.org/abs/1407.5597}{{\ttfamily arXiv:1407.5597
  [hep-th]}}.

\bibitem{Goon:2016mil}
G.~Goon, ``{Heavy Fields and Gravity},''
  \href{http://dx.doi.org/10.1007/JHEP01(2017)045}{{\em JHEP} {\bfseries 01}
  (2017) 045}, \href{http://arxiv.org/abs/1611.02705}{{\ttfamily
  arXiv:1611.02705 [hep-th]}}.
  \href{https://doi.org/10.1007/JHEP03(2017)161}{[Erratum: {\it JHEP} {\bf 03}
  (2017) 161]}.

\bibitem{Note9}
At two loops, $\eta $ is generated even in a purely gravitational theory, with
  a beta function that leads to positive $\eta $ in the asymptotic
  infrared~\cite {Bern:2015xsa,Bern:2017puu}, though this effect is too small
  to be physically relevant, even for astrophysical black holes.

\bibitem{Arkani-Hamed:2006emk}
N.~Arkani-Hamed, L.~Motl, A.~Nicolis, and C.~Vafa, ``{The string landscape,
  black holes and gravity as the weakest force},''
  \href{http://dx.doi.org/10.1088/1126-6708/2007/06/060}{{\em JHEP} {\bfseries
  06} (2007) 060}, \href{http://arxiv.org/abs/hep-th/0601001}{{\ttfamily
  arXiv:hep-th/0601001}}.

\bibitem{Kats:2006xp}
Y.~Kats, L.~Motl, and M.~Padi, ``{Higher-order corrections to mass-charge
  relation of extremal black holes},''
  \href{http://dx.doi.org/10.1088/1126-6708/2007/12/068}{{\em JHEP} {\bfseries
  12} (2007) 068}, \href{http://arxiv.org/abs/hep-th/0606100}{{\ttfamily
  arXiv:hep-th/0606100}}.

\bibitem{Cheung:2019cwi}
C.~Cheung, J.~Liu, and G.~N. Remmen, ``{Entropy Bounds on Effective Field
  Theory from Rotating Dyonic Black Holes},''
  \href{http://dx.doi.org/10.1103/PhysRevD.100.046003}{{\em Phys. Rev. D}
  {\bfseries 100} (2019) 046003},
  \href{http://arxiv.org/abs/1903.09156}{{\ttfamily arXiv:1903.09156
  [hep-th]}}.

\bibitem{Arkani-Hamed:2021ajd}
N.~Arkani-Hamed, Y.-t. Huang, J.-Y. Liu, and G.~N. Remmen, ``{Causality,
  unitarity, and the weak gravity conjecture},''
  \href{http://dx.doi.org/10.1007/JHEP03(2022)083}{{\em JHEP} {\bfseries 03}
  (2022) 083}, \href{http://arxiv.org/abs/2109.13937}{{\ttfamily
  arXiv:2109.13937 [hep-th]}}.

\bibitem{Goon:2019faz}
G.~Goon and R.~Penco, ``{Universal Relation between Corrections to Entropy and
  Extremality},'' \href{http://dx.doi.org/10.1103/PhysRevLett.124.101103}{{\em
  Phys. Rev. Lett.} {\bfseries 124} (2020) 101103},
  \href{http://arxiv.org/abs/1909.05254}{{\ttfamily arXiv:1909.05254
  [hep-th]}}.

\bibitem{Gou:2011nq}
L.~Gou, J.~E. McClintock, M.~J. Reid, J.~A. Orosz, J.~F. Steiner, R.~Narayan,
  J.~Xiang, R.~A. Remillard, K.~A. Arnaud, and S.~W. Davis, ``{The Extreme Spin
  of the Black Hole in Cygnus X-1},''
  \href{http://dx.doi.org/10.1088/0004-637X/742/2/85}{{\em Astrophys. J.}
  {\bfseries 742} (2011) 85}, \href{http://arxiv.org/abs/1106.3690}{{\ttfamily
  arXiv:1106.3690 [astro-ph.HE]}}.

\bibitem{Risaliti:2013cga}
G.~Risaliti { et~al.}, ``{A rapidly spinning supermassive black hole at the
  centre of NGC 1365},'' \href{http://dx.doi.org/10.1038/nature11938}{{\em
  Nature} {\bfseries 494} (2013) 449},
  \href{http://arxiv.org/abs/1302.7002}{{\ttfamily arXiv:1302.7002
  [astro-ph.HE]}}.

\bibitem{Zackay:2019tzo}
B.~Zackay, T.~Venumadhav, L.~Dai, J.~Roulet, and M.~Zaldarriaga, ``{Highly
  spinning and aligned binary black hole merger in the Advanced LIGO first
  observing run},'' \href{http://dx.doi.org/10.1103/PhysRevD.100.023007}{{\em
  Phys. Rev. D} {\bfseries 100} (2019) 023007},
  \href{http://arxiv.org/abs/1902.10331}{{\ttfamily arXiv:1902.10331
  [astro-ph.HE]}}.

\bibitem{Cano:2024bhh}
P.~A. Cano and M.~David, ``{Teukolsky equation for near-extremal black holes
  beyond general relativity: near-horizon analysis},''
  \href{http://arxiv.org/abs/2407.02017}{{\ttfamily arXiv:2407.02017 [gr-qc]}}.

\bibitem{Aretakis:2012ei}
S.~Aretakis, ``{Horizon Instability of Extremal Black Holes},''
  \href{http://dx.doi.org/10.4310/ATMP.2015.v19.n3.a1}{{\em Adv. Theor. Math.
  Phys.} {\bfseries 19} (2015) 507--530},
  \href{http://arxiv.org/abs/1206.6598}{{\ttfamily arXiv:1206.6598 [gr-qc]}}.

\bibitem{Gajic:2023uwh}
D.~Gajic, ``{Azimuthal instabilities on extremal Kerr},''
  \href{http://arxiv.org/abs/2302.06636}{{\ttfamily arXiv:2302.06636 [gr-qc]}}.

\bibitem{Caron-Huot:2022ugt}
S.~Caron-Huot, Y.-Z. Li, J.~Parra-Martinez, and D.~Simmons-Duffin, ``{Causality
  constraints on corrections to Einstein gravity},''
  \href{http://arxiv.org/abs/2201.06602}{{\ttfamily arXiv:2201.06602
  [hep-th]}}.

\bibitem{xAct}
J.~M. Mart\'{i}n-Garc\'{i}a, ``{{\it xAct}: Efficient tensor computer algebra
  for the Wolfram Language}.'' \url{http://www.xact.es/}.

\bibitem{Bern:2015xsa}
Z.~Bern, C.~Cheung, H.-H. Chi, S.~Davies, L.~Dixon, and J.~Nohle, ``{Evanescent
  Effects Can Alter Ultraviolet Divergences in Quantum Gravity without Physical
  Consequences},'' \href{http://dx.doi.org/10.1103/PhysRevLett.115.211301}{{\em
  Phys. Rev. Lett.} {\bfseries 115} (2015) 211301},
  \href{http://arxiv.org/abs/1507.06118}{{\ttfamily arXiv:1507.06118
  [hep-th]}}.

\bibitem{Bern:2017puu}
Z.~Bern, H.-H. Chi, L.~Dixon, and A.~Edison, ``{Two-Loop Renormalization of
  Quantum Gravity Simplified},''
  \href{http://dx.doi.org/10.1103/PhysRevD.95.046013}{{\em Phys. Rev. D}
  {\bfseries 95} (2017) 046013},
  \href{http://arxiv.org/abs/1701.02422}{{\ttfamily arXiv:1701.02422
  [hep-th]}}.

\end{thebibliography}\endgroup
